# Towards Agentic AI Governance: A Preliminary Assessment

Mubarak Raji[0009-0003-0287-4393] and Masooda Bashir [0000-0001-8299-3111]

School of Information Sciences, University of Illinois at Urbana-Champaign, IL, 61820 USA
{mraji2@illinois.edu; mnb@ illinois.edu}

**Abstract.** Artificial intelligence is rapidly evolving from generative systems to agentic AI capable of autonomously planning and executing tasks. Widely characterized as the "Year of Agentic AI," 2025 marked accelerated development and deployment, introducing new ethical and governance challenges. This paper presents a systematic review of the emerging literature on agentic AI governance. Our analysis identifies features that distinguish agentic AI from traditional systems and why it warrants targeted governance attention. We synthesize prevailing governance priorities, proposed mechanisms, and stakeholder roles shaping this evolving domain. As an initial scholarly effort, this review lays the preliminary groundwork for developing a structured roadmap to guide responsible and adaptive agentic AI governance.

**Keywords:** Agentic AI governance, Agentic AI regulation, AI policy, Ethical AI agents.

## 1 Introduction

Artificial Intelligence (AI) has revolutionized our world and the way we do things. We have seen some streams of AI innovations, from traditional AI to generative AI, AI agents, and the new trending AI, agentic AI. Agentic AI (AgenAI ) is an autonomous AI system that can perceive a task, learn and adapt to its environment, and execute the task autonomously to reach a particular goal with little or no human oversight [22]. In other words, agentic AI can accomplish a task on its own on behalf of a human. There have been increases in the adoption of Agentic AI and AI agents by organizations, with about 11% deployment in the first quarter of 2025 and 42% deployment in the third quarter of 2025, as reported by KPMG [19]. In a 2025 PWC survey of business leaders, 88% of top executives surveyed stated that the rise of agentic AI had led to a budget increase on AI products, with about a 66% increase in work productivity for organizations that have adopted the use of AI Agents [30]. Gartner predicts that by 2029, agentic AI, which is classified as the future now, will be the "new normal" and will normalize human agents' collaboration in our day-to-day activities [13]. Apart from adoption, the growth of the agentic AI market is increasing sporadically, with about USD 4.81 billion in 2024 to USD 7.06 billion in 2025, with a market forecast of USD 93.20 billion by 2032, as analyzed by MarketsandMarkets [21].

Agentic AI has many potential benefits, such as increasing work productivity, enhancing economic progress, workplace reshaping and upskilling, advancement of science and technology, faster decision-making process, making essential services more reliable and accessible, among others [31]. Agentic AI can be used across several sectors: financial technology [10], business process management [43], urban planning and smart cities[41], employment, healthcare, customer service, general finance, manufacturing and industry, transportation, energy, cybersecurity, e-commerce, software engineering, among others [1, 12, 15, 29, 33].

While agentic AI has many potentials, there are unique aspects of agentic AI that governance cannot overlook. Some of the unique features of Agentic AI and its possible consequences make governance of agentic AI an important and critical topic, which requires adequate attention, especially since innovation and invention always precede regulation. Prior research has identified ethical concerns, safety, risk, technical vulnerabilities which can make it unpredictable, agency problems in determining the principal, absence of consistent human oversight, uncertainty of regulation and governance approach, efficiency, learning ability, adaptability, value alignment, reliability, scalability among others [20, 31]. These may lead to bias, algorithmic injustice, power imbalance issues, and inequality among others [1, 36].

Since agentic AI is also part of the larger AI family, ethical AI governance principles and data privacy frameworks are also applicable to it; however, the governance approach may differ in some instances due to agentic AI's unique peculiarities. For example, breach of data privacy laws can expose violators to different kinds of enforcement, including administrative, civil, criminal, or financial sanctions (Raji et al., 2025) [32]. Governance of Agentic AI is also critical when minimizing the effect of surveillance capitalism [45], data colonialism [9], digital colonialism [8], racial discrimination and bias, algorithm injustice, or profiling, especially against minority groups.

Despite the distinctive capabilities of agentic AI and the corresponding need for tailored governance mechanisms, no dedicated governance framework currently exists beyond Singapore's *Model AI Governance Framework for Agentic AI*, released at the World Economic Forum in Davos on January 22, 2026. To date, this appears to be the only formal framework explicitly addressing agentic AI. Notably, neither the European Union, whose regulatory instruments such as the GDPR have shaped global technology governance, nor the United States has introduced a specific governance framework for agentic AI. In response to this gap, this paper presents a foundational scholarly effort to advance the development of an agentic AI governance framework. Through a systematic review of existing AI governance literature, we identify the distinctive features and risk dimensions of agentic AI that warrant dedicated consideration and propose an initial roadmap to inform future governance design.

To situate our contribution, Section 2 provides a concise background on the evolution of agentic AI and the role of governance, as well as related work on AI governance. Section 3 details our systematic review methodology, including the criteria for paper selection and the thematic analysis used to identify governance-relevant patterns in the literature. We then present emerging issues in agentic AI governance, synthesizing key findings across peer-reviewed studies, including definitional and classification challenges (e.g., agentic AI as autonomous pursuit systems and questions of moral agency), privacy implications, stakeholder roles, and the relevance of Singapore's Model AI Governance Framework for Agentic AI. The paper concludes with a discussion of limitations and a forward-looking research agenda to guide the continued development of agentic AI governance.

## 2 Background

### 2.1 From Traditional AI to Agentic AI

The history of artificial intelligence dates back to the 1950s when John McCarthy devised the concept of AI [40, 44]. This was the beginning of the now popular innovation that everybody uses in their daily lives, work, and day-to-day activities. AI has many developmental stages, some of which are its downtime, even though AI research was ongoing. It became more popular with a major comeback when ChatGPT was released to the public in 2021[40]. Hosseini & Seilani captured the evolution of artificial intelligence from its early beginning to the current stage into five categories; we add the sixth stage, which is the current stage. The first stage is before 2000, where problems are solved using "hand-crafted rules," which require express instructions or are manually programmed and are non-adaptive in nature [15]. The second stage post 2000 was the adoption of basic Natural Language Processing (NLP) techniques using statistical models for learning, and the system can interact and perform from available data [15]. The third stage, around 2010, birthed deep learning, which enables systems to work more efficiently through innovations such as "Convolutional Neural Networks (CNNs)" and the "transformer architecture," making interaction with multimodal inputs such as audio, images, and text better [15]. The fourth stage, which was post-2010 but before 2020, was indeed the gamechanger, which saw the invention of generative AI in different models with better alignment and reasoning. This period gave birth to the "GPT models," which can be used to generate content using prompt engineering [15, 40]. The fifth stage is post 2020, which have seen advance in generative AI, the invention of advance autonomous AI agent, which can be used for instantaneous interaction and capable of perceiving, learning on its own and making decision autonomously [15]. The last stage in our view is 2024 and beyond, which is seeing the rise of agentic AI, which can achieve goal-oriented tasks autonomously with minimal or no human oversight, and the transformation of robotics into physical AI.

### 2.2 Generative AI, AI Agents and Agentic AI

As earlier stated, agentic AI is a special kind of AI systems or tools that autonomously accomplish various tasks on its own using several algorithmic designs and interaction with multiagent to attain complex goals with little or no human involvement or constant supervision. These systems perceive and continuously interacting, adapting, learning, reasoning from their environment to produce required result over a period of time [1, 3, 10, 12, 15, 16, 22, 29, 31, 37]. The agentic loop addresses complexity through perception, reasoning, action, and iterative learning [22, 31]. In other words, agentic AI architecture distill data from multiple sources, observe and recognize information patterns; strategically synthesis those data through retrieval-augmented generation for LLM-driven reasoning for a logical conclusion; execute predefined tasks based on user's input or its reasoning and interact with physical and digital world through API integrations within its guardrail to ensure reliability; and continuously enhance its performance through feedback flywheel or loop [22, 31]. Due to its autonomous goal-pursuit features, agentic AI differs from generative AI (GenAI) and traditional AI tools that require human prompts and specific instructions before carrying out a task. Agentic AI is proactive and utilizes generative models to convert high-level goals into manageable and actionable steps to perform and achieve a goal without constant human involvement. GenAI is reactive, waiting for a prompt to create text, images, or code [1, 10, 15]. Hence, agentic AI is about outcomes (how) while GenAI focuses on artifacts (what). For example, GenAI can help with travel planning and come up with an itinerary without helping to do booking; however, agentic AI can come up with an itinerary, carry out booking, and ensure it does not conflict with other schedules.

### 2.3 Role of AI Governance

Furthermore, with the global resurgence of AI, data is an important commodity for training AI modules. The ethical concerns, such as privacy risks, human control, accountability, and transparency issues of AI pose to humanity, thereby making AI governance a necessity to ensure ethical development and deployment of AI [7]. Similarly, several countries, customs unions, or organizations are coming up with AI governance frameworks [38].

During the stages of AI development explained above, particularly stages four and five, AI governance, ethical principles and policy have also evolved in the form of statutes, guidelines, frameworks, declarations and standards. These includes European Union Artificial Intelligence Act, Colorado AI Act, Council of Europe's Framework Convention on Artificial Intelligence and Human Rights, Democracy and the Rule of Law, United Nations Educational, Scientific and Cultural Organization's (UNESCO) Recommendation on the Ethics of Artificial Intelligence, Organization for Economic Co-operation and Development (OECD) AI Principles, United States Executive Order on the Safe, Secure, and Trustworthy Development and Use of Artificial Intelligence (now abolished), Africa Union Continental Artificial Intelligence Strategy, and National Institute of Standards and Technology (NIST) Artificial Intelligence Risk Management Framework (AI RMF 1.0) among others.

These AI ethical principles include transparency, validity, reliability, explainability, security, interpretability, value alignment, safety, accountability, human agency and oversight, fairness and non-discrimination, sustainability, data privacy, human oversight, diversity and inclusion, proportionality, among others [14]. It is important to note that these ethical AI principles are not contained in a single law or framework [14]. For example, UNESCO addressed proportionality in its AI ethics guidelines, whereas the OECD AI Principles did not.

Similarly, existing data privacy laws and regulations such as the EU General Data Protection Regulation (GDPR), Health Insurance Portability and Accountability Act (HIPAA), California Consumer Privacy Act, and the Nigerian Data Protection Act are still applicable to AI tools where personal data or personally identifiable information are involved.

In this paper, the authors examine how researchers describe the governance of agentic AI through a lens analysis of prior literature.

### 2.4 Related Work

To the best of our knowledge, there has not been any work on any literature review on the governance of agentic AI. However, we noted that there are literature reviews conducted on AI governance generally, which we deem it fit to mention. One of them is Batool et al. (2025), who conducted a systematic review of scholarly literature on AI governance, focusing on who is governing AI, what, when, and how AI is being governed [2]. Similarly, Papagiannidis et al. (2025) did a scoping review to synthesize prior work on responsible AI governance and came up with a framework for organizational practices that include "structural, relational and procedural practices" [28]. Birkstedt et al. (2025) also conducted a systematic literature review to integrate existing studies within the context of the organizational level [4]. They conceptualized organizational-level AI governance, identified trending research areas, and suggested future directions [4]. While these scholarly publications focus on a review of AI governance literature, they are generally broad or centered on organizational practices and do not cover topics relating to AI agents or agentic AI, which is an emerging area, and research on its governance approaches is still at the elementary stage. Our work seeks to fill this gap by looking at the common trends and patterns in scholarly work that are related to the governance of agentic AI.

## 3 Method

The research process started with an extensive search across multiple digital repositories, including Google Scholar, the ACM Digital Library, SSRN, JSTOR, and the AAAI repositories. We focused on literature published from 2020 to 2025, a period chosen because of the novelty of agentic AI as a distinct research area and the recent development of governance frameworks. Searches used key terms like "agentic AI governance," "governance of agentic AI," and "governance of autonomous agentic AI" to capture the latest scholarly discussions.

The initial broad search returned over 3,000 papers from various fields. However, many did not meet our criteria as they mainly contained general AI keywords without a focus on agentic AI governance. We narrowed our scope to papers specifically about AI governance, which identified about 995 relevant works.

To ensure data quality and prevent duplication, we merged duplicates from different databases, resulting in a unique set of papers tailored to our study. We also excluded studies on physical autonomous devices such as autonomous vehicles, military equipment, and drones, as their governance differs from that of software-based AI. This excluded 552 papers that focused on "autonomous" devices but did not address agentic systems.

Further refinement involved removing 389 papers that focused on related but different AI technologies. Research on generative AI, conversational AI, and traditional AI was excluded to maintain a strict focus on agentic AI, its ethical issues, and liability concerns.

This careful filtering led to a final collection of 54 core publications. To ensure academic rigor, we included only works that had undergone peer review [17]. As a result, 33 papers and reports were excluded due to the lack of formal peer review. These were excluded to prioritize peer-reviewed scholarly articles for validity and reliability [17].

After excluding non peer-reviewed work, the final set of 21 articles listed below (Table1) was selected and reviewed for this study.

Upon identifying our literature, we created a table to identify criteria for thematic analysis to see the common themes and patterns on agentic AI governance in the literature.

**Table 1.** List of scholarly papers considered in this study.

| Authors/ Discipline | Year | Title | Venue | Core Methodology |
|---|---|---|---|---|
| Acharya et al. [1] Interdisciplinary | 2025 | Agentic AI: Autonomous intelligence for complex goals–a comprehensive survey | IEEE Access | Conceptual/Theoretical |
| Bent [3] Engineering | 2025 | The Term "Agent" Has Been Diluted Beyond Utility and Requires Redefinition | AIES[1] | Conceptual/Theoretical Review |
| Chan et al. [6] Interdisciplinary | 2024 | Visibility into AI Agents | FAccT[2] | Conceptual/Theoretical Technical Proposal |
| Chan et al. [5] Interdisciplinary | 2023 | Harms from increasingly agentic algorithmic systems. | FAccT | Conceptual/Theoretical Technical Proposal |
| Dodda [10] Computer Science (CS) | 2023 | AI Governance and Security in Fintech: Ensuring Trust in Generative and Agentic AI Systems | AAJED[3] | Conceptual/Theoretical Regulatory Landscape Case study |
| Gahnberg [11] Political Science & International Relations | 2021 | What rules? Framing the governance of artificial agency | Policy Studies Journal | Conceptual/Theoretical |
| Garg [12] Healthcare Innovation | 2025 | Designing the Mind: How Agentic Frameworks Are Shaping the Future of AI Behavior | Journal of Computer Science and Technology Studies | Conceptual/Theoretical |
| Hosseini & Seilani [15] CS / Engineering | 2025 | The Role of Agentic AI in Shaping a Smart Future: A Systematic Review | Array | Conceptual/Theoretical Review |
| Hughes et al. [16] Interdisciplinary | 2025 | AI agents and agentic systems: A multi-expert analysis | Journal of Computer Information Systems | Conceptual/Theoretical Review |

[1] AIES stands for AAAI/ACM Conference on AI, Ethics, and Society.
[2] FAccT stands for ACM Conference on Fairness, Accountability, and Transparency.
[3] AAJED stands for American Advanced Journal for Emerging Disciplinaries.

| | | | | |
|---|---|---|---|---|
| Kolt [18] Law & CS | 2025 | Governing AI Agents | Notre Dame Law Review | Conceptual/Theoretical Legal Analysis |
| Lior [20] Law | 2020 | AI Entities as AI Agents: Artificial Intelligence Liability and the AI Respondeat Superior Analogy | Mitchell Hamline Law Review | Legal Analysis |
| Murugesan [22] Engineering | 2025 | The rise of agentic AI: implications, concerns, and the path forward | IEEE Intelligent Systems | Conceptual/Theoretical |
| Nasim [23] Math & Physics | 2025 | Governance in agentic workflows: Leveraging LLMs as oversight agents. | AAAI 2025 Workshop on AI Governance | Technical Proposal Conceptual/Theoretical |
| Navaie [24] CS | 2025 | From rights to runtime: Privacy engineering for agentic AI | AI Magazine (by AAAI) | Technical Proposal Regulatory Landscape |
| O'Keefe et al. Law[25] | 2025 | Law-Following AI: Designing AI Agents to Obey Human Laws | Fordham Law Review | Legal Analysis |
| Pawar [31] CS | 2025 | Ethical and Governance Challenges of Agentic AI | International Journal of Humanities and Information Technology | Qualitative |
| Raheem & Hossain [31] CS | 2025 | Agentic AI Systems: Opportunities, Challenges, and Trustworthiness | IEEE International Conference on Electro Information Technology | Conceptual/Theoretical |
| Riedl & Desai [33] CS & Law | 2025 | AI Agents and the Law | AIES | Legal Analysis Conceptual/Theoretical |
| Saloustrou [35] | 2025 | Agentic AI Regulation under the AI Act: A Proof of Concept or a Concept to Prove | Journal of AI Law and Regulation | Legal Analysis Regulatory Landscape |
| Tiwari [41] Urban and Regional Planning | 2025 | Conceptualising the emergence of Agentic Urban AI: from automation to agency | Urban Informatics (by Springer) | Conceptual/Theoretical |
| Vu et al. [43] Interdisciplinary | 2025 | Agentic Business Process Management: Practitioner Perspectives on Agent Governance in Business Processes | International Conference on Business Process Management | Empirical/ Qualitative |

## 4 Developing Issues on Agentic AI Governance

In this section, we present the key findings from our review of the selected literature, highlighting the recurring themes and governance patterns identified across publications on agentic AI.

## 4.1 Definitions and Classification Syndrome

As highlighted by Bent, the definition and classification of agentic AI are the foundation and fulcrum of effective agentic AI governance, as misdirection in the definition may lead to wrong governance or policy approach, as it may mislead policymakers [3]. Gahnberg also observed that governing AI involves understanding artificial agents and their material agency and components [11]. While Navaie acknowledges that there has not been a clear legal definition of agentic AI as of the end of 2025 [24], the Singapore Agentic AI MGF provides a working definition, even though the framework still lacks a generally acceptable definition of what constitutes an agent. This agentic AI definition syndrome can be akin to Professor Solove's description of privacy as "a concept in disarray" in his award-winning work, "Taxonomy of Privacy" [39]. Singapore Agentic AI Framework defined agentic AI as "systems that can plan across multiple steps to achieve specified objectives, using AI agents. There is no consensus on what defines an agent, but there are certain common features – agents usually possess some degree of independent planning and action taking (e.g., searching the web or creating files) over multiple steps to achieve a user-defined goal." This definition takes an autonomous goal-pursuit approach, which will be discussed below.

We observed two major approaches to classifying or discussing agentic AI governance: autonomous goal-pursuit and tool, which describe agentic AI as an autonomous AI system that can be used to achieve a particular goal and as a tool for planning and carrying out tasks; moral agency, as an agent of a particular principal.

### A. Autonomous goal-pursuit

This classification focuses on the ability and capacity to complete a complex task independently to achieve a particular goal by learning from the environment. We observe that many authors such as Chan et al. (2023), Bent and Gahnberg among others, relied on the outstanding work of Russell and Norvig on autonomy of an agent that, "to the extent that an agent relies on the prior knowledge of its designer rather than on its own percepts and learning processes, we say that the agent lacks autonomy" [34, 60]. This means that an agent is considered to lack autonomy when it depends more on the knowledge and insights provided by its designer than on its own observations and learning experiences. Based on the 21 scholarly works considered, we curate a list of features or attributes of an agentic AI in Table 2 below.

**Table 2.** Attributes of Agentic AI.

| Features | Definitions | Literature |
|---|---|---|
| Adaptability | Agility and capacity to react environmental shifts efficiently | Acharya et al., Bent, Garg, Hosseini & Seilani, Hughes et al., Murugesan, Nasim, Navaie, O'Keefe et al., Pawar, Raheem & Hossain, Saloustrou, Tiwari, Vu et al. |
| Autonomy | Ability to perform, perceive, predict, learn, plan, and carry out a task independently without or little human intervention | Acharya et al., Bent, Gahnberg, Garg, Hosseini & Seilani, Hughes et al., Murugesan, Nasim, Navaie, O'Keefe et al., Pawar, Raheem & Hossain, Saloustrou, Tiwari, Vu et al. |
| Goal complexity | Agentic AI are engineered to target definite complex goals, and their performance are streamlined to accomplish the expected objectives. | Acharya et al., Bent, Chan et al. (2023), Chan et al. (2024), Gahnberg, Garg, Hosseini & Seilani, Hughes et al., Murugesan, Nasim, Pawar, O'Keefe et al., Raheem & Hossain, Saloustrou, Tiwari, Vu et al. |
| Environmental interaction | Capacity to engage their environment, detecting shifts, sense alterations and modify their approaches to perform optimally in intricate situations | Acharya et al., Bent, Gahnberg, Garg, Hosseini & Seilani, Hughes et al., Murugesan, Nasim, Navaie, O'Keefe et al., Raheem & Hossain, Saloustrou, Tiwari, Vu et al. |
| Learning capability | Ability to learn on itself from available data (self-learning) and through iterative process or trial and error (reinforcement learning) when interacting with its environment over a period of time. | Acharya et al., Bent, Gahnberg, Garg, Hosseini & Seilani, Hughes et al., Murugesan, Nasim, O'Keefe et al., Pawar, Raheem & Hossain, Saloustrou, Vu et al. |
| Workflow optimization | Ability carry out work efficiently by streamlining workflows using its core abilities | Garg, Hosseini & Seilani, Hughes et al., Murugesan, O'Keefe et al., Pawar, Raheem & Hossain, Saloustrou, Vu et al. |
| Multi-agent systems | Capability to dialogue and coordinate with different agents and systems in establishing pipelines to execute broad spectrum of tasks and responsibilities | Acharya et al., Garg, Hosseini & Seilani, Hughes et al., Murugesan, Nasim, Raheem & Hossain, Saloustrou, Vu et al. |
| Temporal Coherence | Capacity to uphold consistent function across periods of time by utilizing its awareness of its current state and its memory | Bent |

We observed that of all the attributes listed above, "temporal coherence" is unique and was recently coined by Bent in her AIES 2025 paper, and it relates to the memory capacity of agentic AI [3]. However, other authors, directly or through closely related features, discussed agentic AI memory capacity in relation to long-term memory objectives and the ability to learn from past interactions. Another critical observation is that some authors did not expressly list some of the attributes above in their work; the feature can still be inferred from the collective reading of their work. For example, Acharya et al did not expressly mention workflow optimization but identified that agentic AI plays task management functions [1].

### B. Moral Agency

Another lens through which agency is discussed is through the lens of moral agency, which involves the ability to make autonomous decisions and who is responsible or will be held accountable for those decisions as an agent. This is mostly discussed using the common law approach of the principal and agent relationship. This approach is championed by Kolt [18], Lior [20] and O'Keefe et al [25] from a legal perspective and Riedl & Desai [33], from a techno-legal perspective. Under agency theory, an agent carries out an instruction on behalf of a disclosed principal through delegation, which is binding on the principal, and the agent holds a fiduciary duty of loyalty to the principal [18, 20, 25, 33]. The principal instruction can be carried out through actual authority, express instruction, or implied authority, which can be inferred from the circumstances [33]. However, some authors argued that current agency law theory needs some modifications to adequately address principal and agents relationship of AI agents because of arising problems [18, 20, 25, 33]. Kolt identified that information asymmetry affects disclosure of material facts as it is difficult to determine what the agents know and what they ought to know; determining the scope of authority can be unclear to ascertain the principal's manifestation and extent of the agent's discretion; and AI agents may breach loyalty duty when it fails to act in the principal's best interest [18]. He further argued that the incentive approach used in rewarding and penalizing, oversight and enforcement methods deployed in human agency, may not be applicable to AI agents as they operate and make a case for governance with inclusive values, clear visibility into operations, and distributed liability for harms to ensure safety and ethics [18].

Lior, in her earlier work, raises the concern of how to identify a principal, which is crucial in determining liability. She stated that a principal can be identified by looking at level of involvement, supervisory power and ability to directly control the AI agent's actions [20]. Using the respondeat superior principle, which makes a principal liable for an agent's conduct, Lior argues that a strict liability approach should be deployed in cases of harm using several legal analogies: place a major liability burden on product manufacturers, and hackers when there is a security breach; and acknowledge there can be multiple principals (developer, deployer) as well as joint liability. She calls for an amendment of agency laws to recognize computer programs as agents [20].

Riedl & Desai argued that AI agents necessitate expanded value alignment beyond helpfulness, honesty, and harmlessness. To mitigate third-party harms and foster trust, alignment must explicitly incorporate principles of disclosure and loyalty. This proactive approach guides responsible agent behavior and builds confidence in their real-world interactions [33].

O'Keefe et al. propose designing AI agents as "Law-Following AIs" (LFAI), focusing on legal compliance and refusal of illegal actions, even if directed by their principals [25]. They see this legal obligation as a necessary evolution in legal ontology, enabled by AI's ability to reason about laws [25]. These AI agents are considered "legal actors" capable of action and duties, distinct from "legal persons" [25]. This approach helps agents follow legal duties over primary instructions, prevents de facto actors without de jure duties, and promotes accountability, avoiding "AI henchmen" [25].

Hughes et al. make a case for recognition of limited legal personhood for agentic AI to allow them to enter into contracts and assume legal responsibility for their conduct [16]. However, Lior rejected this approach, arguing that treating AI as an electronic person raises ethical issues and its developer or maker should be held accountable for its damages [20]. It is important to observe that these approaches rely heavily on the common law approach of the principal-agent relationship and exclude other legal systems around the world, such as civil law, international law, religious laws, or customary laws.

### 4.2 Privacy

Privacy has always been a germane issue to any innovation that uses personal data. Agentic AI features give room for self-learning and reinforcement learning based on the data gathered from its environment. Temporal coherence also enables agentic AI to store large amounts of data in memory for a period to meet specific operational requirements. The storage, reuse, and extent of data collection raise a lot of privacy concerns relating to the principles, purpose limitation, data minimization, storage limitation, automated decision making, transparency, accountability, right of erasure, among others, especially as privacy laws mostly focus on data collection and leave repurposed data.

Navaie argued that existing privacy laws can be used to address agentic AI governance issues and develop a toolkit for privacy engineers using the GDPR, CPRA, LGPD, and PDPA as case studies [24]. These underscore that existing laws can be repurposed to address challenges of an upcoming innovation, and they will serve as a practical guide for privacy professionals. While this will serve as an interim measure, Solove argued that it may not adequately address the privacy issues arising from the development and deployment of AI [40], just like many data privacy laws have not addressed the problem arising from data brokage and surveillance capitalism.

### 4.3 Singapore Model AI Governance Framework for Agentic AI

As stated earlier, Singapore Agentic AI MGF was unveiled in January 2026 to specifically address and serve as a governance model for agentic AI [46]. This is the first governance model in the world specifically on agentic AI to help organizations with a governance model while developing or deploying agentic AI internally or outsourcing from a third-party supplier to optimize workflow. It defined agentic AI as autonomous goal-pursuit, as discussed in 4.1 above. It addresses its core components, the sources and types of agentic AI risk, and recommends four dimensions of governing agentic AI [46]. One crucial part of the Singapore Agentic AI MGF is that human oversight is an integral part of the governance of agentic AI and must be held accountable for risks associated with agentic AI, which is in line with Lior's argument that the AI agents do not need "electronic personality" [20, 46].

The types of risk enumerated in the model include “erroneous actions,” “unauthorized actions,” “biased or unfair actions,” “data breaches” and “disruption to connected systems.” It is important to note that the Singapore Agentic AI MGF does not classify risk in the same manner as the EU AI Act classifies risk for general-purpose AI, which includes unacceptable, high, limited, and minimal risk. Saloustrou posits that the EU AI Act can serve as a foundational “proof of concept” for agentic AI governance, demonstrating an initial approach to governance. However, its effectiveness remains unproven due to several limitations [35]. Saloustrou also argued that the EU AI Act faces significant difficulties in addressing the high autonomy of agentic AI, which can make decisions independently beyond predefined parameters . Additionally, she further stated that the EU Act suffers from definitional ambiguities, such as vague criteria for what constitutes an AI agent, and oversight challenges, including monitoring complex autonomous behaviors [27, 35]. These issues collectively result in a regulatory structure that may be insufficient to adequately manage the risks associated with advanced agentic AI [35].

Singapore Agentic AI MGF risk classification focuses on types of negative consequences arising from agent malfunction. The four dimensions in the Singapore Agentic AI MGF of governance are: “assess and bound the risks upfront,” “make humans meaningfully accountable,” “implement technical controls and processes,” and ‘enable end-user responsibility.”

### 4.4 Role of stakeholders and why their participation is critical

In this section, we identify several key stakeholders that can play a crucial role in leading the governance of agentic AI. They include international organizations such as the OECD to come up with a governance framework just like they did with the OECD AI governance principles and the OECD Privacy Principles (1980/ 2013), which can serve as a model for their member countries. In addition, United Nations could be a key figure in global governance of Agentic AI as they are already leading many initiatives related to global governance of AI [42], as their global reach will help in understanding the peculiarities of various communities, especially less-developed countries around the world. Regulators as well as policymakers also have an important role to play; they can work together to come up with policies and frameworks that can adequately address issues of risk classifications, periodic audits, and risk impact assessments .

Developers who play the role of manufacturer have a role to play in incorporating ethics by design, just like privacy by design, at the earliest stage of developing agentic AI systems to ensure accountability, foster transparency, and eliminate bias [1, 5, 6, 12, 15, 37]. Additionally, until there is a specific governance approach that expressly clarifies the liability of each stakeholder, as Lior posits, the major responsibility is still on the developer [20]. The deployer also has a role to play as they will ensure constant oversight and oversee constant human-AI collaboration and need to conduct constant audits or impact assessments [23, 37].

For researchers, whether in the academic community or industry, they have a role to play in turning their research into a practical governance approach, and a multidisciplinary approach can produce a socio-technical-legal approach that can be adequate [3]. It is also good for agentic AI users to give persistent feedback to agentic AI developers so that they can improve the usability and encourage ethics by design.

Finally, one major aspect that occurred in the early days of AI governance is that governance approach is focus on Global North to the exclusion of the Global Majority, to avoid a repeat of this, countries, researchers, policymakers, governance experts in the Global Majority should also get themselves involved on talks, discussion, research on how to improve or contribute to the governance of agentic AI to ensure their unique peculiarities are addressed [26].

## 5 Limitation and Future Research

One major limitation is that our review focuses mostly on peer-reviewed scholarly work, as explained in our method section above, which was deployed to maintain academic rigor. Additionally, we observed that the discussion of moral agency theory focuses on common law doctrine and excludes other legal systems, such as civil law commonly used in Europe, Latin America, and Francophone African countries, as well as religious and customary laws. This may limit the perspective on moral agency theory to common law, and future research can explore agency theory from the angle of these legal systems.

Additionally, with the growing adoption of agentic AI, empirical researchers can focus on the governance approach of developers of agentic AI to examine whether ethics by design has been deployed. Research also on privacy policies of agentic AI tools to examine how personal data are being collected, used, stored, and deleted.

## 6 Conclusion

Governance of agentic AI is still at an early stage because much of the research and innovation efforts are currently prioritizing the development and deployment of agentic AI systems. Understanding the attributes and definition of agentic AI is very important and an initial step to governance. We observed two major classification agents of AI: first, as an autonomous goal pursuit AI system that has features such as adaptability, autonomy, goal complexity, environmental interaction, learning capability, workflow optimization, and multi-agent systems. The second approach is through the lens of moral agency theory, which was discussed based on the principal-agent relationship. Privacy is still a major concern, and the current privacy laws serve as an interim measure to address privacy-related issues with agentic AI, but do not take the place of a new governance model for agentic AI. Singapore Agentic AI MGF is the first agentic AI governance model in the world. While this is a good initiative, we call for the inclusion of relevant stakeholders in the governance approach to agentic AI.

**Acknowledgments.** We thank the reviewers and participants at AIR-RES 2026 for their comments on the draft paper and presentation.

**Disclosure.** Grammarly was used for language clarity and grammar purposes. The authors assume all responsibility for the content of this submission.